\documentclass{sig-alternate}

\usepackage{graphics}
\usepackage{graphicx}
\usepackage{ifpdf}
\ifpdf
\usepackage[pdftex,pdfdisplaydoctitle,pdfpagemode=None]{hyperref}
\else
\usepackage{hyperref}
\fi

\ifpdf
\hypersetup{
  pdftitle={Recommending Related Papers Based on Digital Library Access Records},
  pdfauthor={Stefan Pohl, Filip Radlinski and Thorsten Joachims},
  pdfkeywords={Recommendations, co-citation, co-download, http access logs}
}
\fi

\makeatletter
\def\section{%
    \@startsection{section}{1}{\z@}{-9\p@ \@plus -4\p@ \@minus -2\p@}
    {4\p@}{\baselineskip 10pt\secfnt\@ucheadtrue}%
}
\def\subsection{%
    \@startsection{subsection}{2}{\z@}{-7\p@ \@plus -2\p@ \@minus -\p@}
    {4\p@}{\secfnt}%
}
\makeatother

\begin{document}

\setlength{\textfloatsep}{0.4cm} 
\setlength{\floatsep}{0.3cm} 

\conferenceinfo{JCDL'07,} {June 17--22, 2007, Vancouver, British Columbia, Canada.}
\CopyrightYear{2007}
\crdata{978-1-59593-644-8/07/0006} 

\title{Recommending Related Papers\\Based on Digital Library Access Records\vspace{-0.65cm}}

\numberofauthors{3}
\author{
\alignauthor Stefan Pohl \\
       \affaddr{Cornell University}\\
       \affaddr{Ithaca, NY, USA}\\
       \email{sp424@cs.cornell.edu}
\alignauthor Filip Radlinski\\
       \affaddr{Cornell University}\\
       \affaddr{Ithaca, NY, USA}\\
       \email{filip@cs.cornell.edu}
\alignauthor Thorsten Joachims\\
       \affaddr{Cornell University}\\
       \affaddr{Ithaca, NY, USA}\\
       \email{tj@cs.cornell.edu}
}

\maketitle
\begin{abstract} 
An important goal for digital libraries is to enable researchers to more easily explore related work. While citation data is often used as an indicator of relatedness, in this paper we demonstrate that digital access records (e.g.\ http-server logs) can be used as indicators as well. In particular, we show that measures based on co-access provide better coverage than co-citation, that they are available much sooner, and that they are more accurate for recent papers.
\end{abstract}

\category{H.3.7}{Information Storage and Retrieval}{Digital Libraries}
\category{H.3.3}{Information Storage and Retrieval}{Information Search and Retrieval\vspace{-1mm}}

\section*{General Terms: \textrm{Algorithms, Experimentation}\vspace{0mm}}
\section*{Keywords: \textrm{Recommendations, co-citation, co-download, http access logs}\vspace{2mm}}

\section{Introduction}

In scientific literature, citation information is a key source of information about relationships between documents. Citations are used to measure impact of documents and journals \cite{Garfield72_citation}, to identify related papers via co-citation and bibliographic coupling \cite{Small73_cocitation, McNee2002}, and to improve ranking in keyword-based search \cite{Page98_pagerank}. Unfortunately, there are at least two problems with citation data. First, extracting citations from academic articles requires manual curation or sophisticated natural language processing. This makes such data costly and time-consuming to obtain. Second, it takes considerable time for a newly published article to gather a sufficient number of citations for meaningful statistical analysis.

We demonstrate that access data of the form ``user X downloaded document Y'' can be used as a substitute for citation data. Access data does not suffer from the drawbacks of citation data, since it is available sooner and can easily be extracted from digital library access logs. 

In this paper, we focus on using access data to identify related papers. Treating access data as a bi-partite graph of users and documents analogous to item-to-item recommendation systems (see e.g.\ \cite{Sarwar/etal/01}), we explore an access-based measure to quantify the degree to which pairs of articles are related. We evaluate how well this measure predicts future co-citations on the arXiv e-Print archive \cite{arxiv}. Our results show that access-based measures have vastly larger coverage and are more accurate at finding related work than co-citation for recently published papers. Additional and more detailed results can be found in \cite{Pohl06}.

\section{Access Data}

The arXiv collection recorded over 650 million accesses to over 350,000 scientific documents between 1994 and July 2006. For each access, we extracted the time and date, source IP address and document accessed. After filtering proxies and crawlers, we segmented accesses into 30 minute sessions from each IP, assuming these to define individual users. This gives, for each session, a set of documents downloaded by the user. Analogous to the co-citation measure of relatedness \cite{Small73_cocitation, McNee2002}, we then transform these sets into counts of how often each pair of documents was co-downloaded.

When dealing with access data, care has to be taken to avoid presentation biases \cite{Joachims/etal/05a}. In particular, we found that access data is influenced by publication date. First, older papers tend to be less accessed. Second, papers published at the same time are often presented on the same web page and tend to be co-accessed more often. For arXiv, this is especially visible during the first month after publication: Many users subscribe to announcements listing all new articles. In contrast, we expect most valuable access data to result from searches of users for a specific topic. Hence we ignored co-downloads of documents appearing together that occurred during the first month after publication.

\section{Evaluation}

We compare co-citation and co-download in terms of coverage and recommendation quality.

\subsection{Coverage of Co-access Data}

\begin{figure}
\centering
\includegraphics[width=0.9\columnwidth]{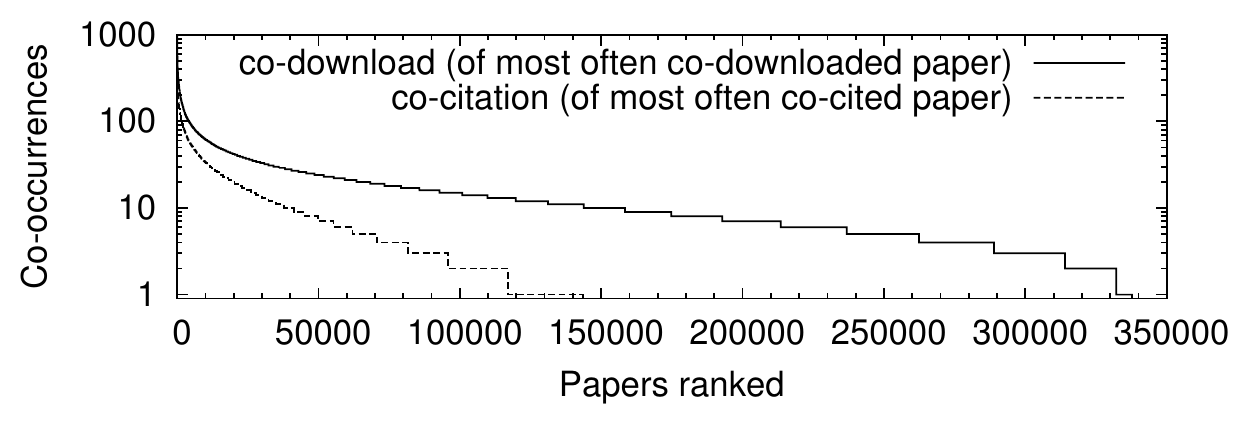}
\vspace{-0.4cm}
\caption{Co-citation and co-download coverage}\label{fig:cocitacccov}
\vspace{0.1cm}
\end{figure}

Figure~\ref{fig:cocitacccov} shows the maximum number of times each paper in arXiv was co-cited and co-downloaded with any one other paper. The papers are sorted by this count independently along the horizontal axis. We see that using co-download, for almost every paper in arXiv we have data and are therefore able to make recommendations. In contrast, about two thirds of the papers have no known co-citations, and those that do often are co-cited only once or twice. Hence co-citation cannot make any recommendations for most papers.

\begin{figure}
\centering
\includegraphics[width=0.9\columnwidth]{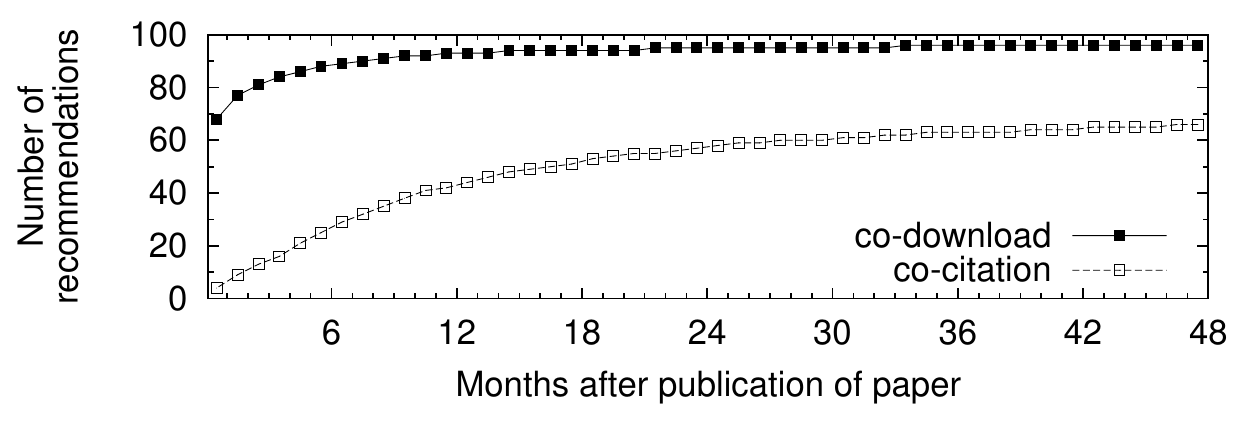}
\vspace{-0.4cm}
\caption{Count of recommendations over paper age}\label{fig:eval3_1}
\end{figure}

More detail about the number of recommendations that co-download and co-citation can make is given in Figure~\ref{fig:eval3_1}. The graph shows the mean number of recommendations (i.e.\ papers with non-zero co-downloads or co-citations) as a function of time after publication of a paper. For efficiency, we limit ourselves to 100 recommendations per paper. We see that the number of recommendations from co-downloads quickly reaches the maximum. In contrast, the average number of recommendations from co-citations grows much more slowly.
Thus, building a citation based recommendation system takes considerably more time than using access data.

\subsection{Recommendation Quality}

While Figure~\ref{fig:cocitacccov} shows that the number of co-down\-loads is much larger than the number of co-citations, the former might be more noisy.
An author's decision to cite a paper is likely to mean more than a download by an anonymous user, who might never even look at the document.
To examine if repeated measurement compensates for such noise, we now evaluate the quality of recommendations.

We use the number of co-downloads before 2005 as a measure of similarity between two papers. In particular, for a given paper, we recommend related papers by sorting all other papers by this similarity. As ground truth, we took the publications after 2005 and assumed that papers cited together are related, and all other papers are unrelated. The citation data came from manual processing of roughly 200,000 arXiv submissions administered in the SLAC/SPIRES database.
To estimate the quality of the recommendations, we take one paper D from a set of references in a 2005 paper, calculate the Mean Average Precision (MAP) \cite{baezayates99modern} for the recommendations made by co-download, and aggregate the MAP scores as a function of the time after publication of D. We report average results over 7,500 papers. We also performed the same experiment using co-citation instead of co-download.

\begin{figure}
\centering
\includegraphics[width=0.9\columnwidth]{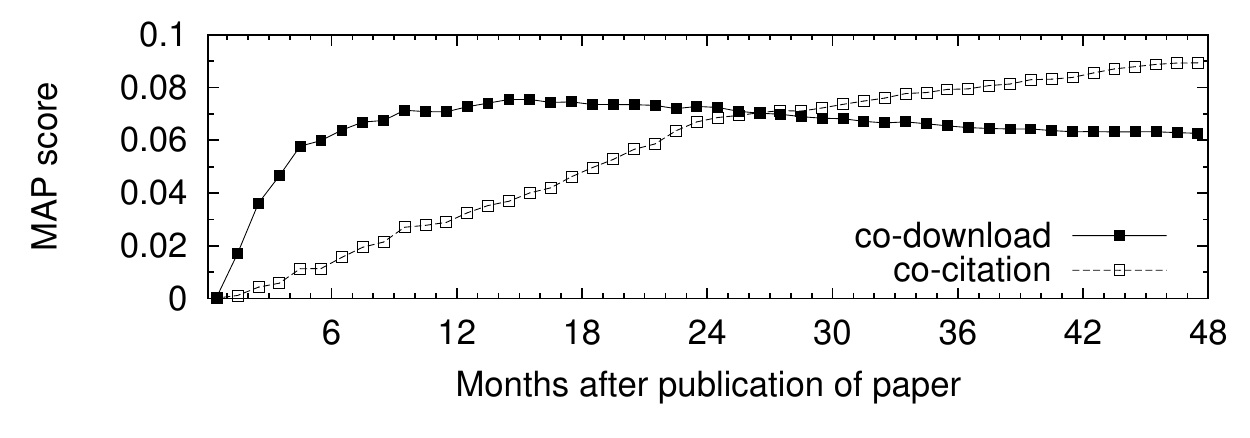}
\vspace{-0.4cm}
\caption{Mean average precision over paper age}\label{fig:eval3_2}
\end{figure}

Figure~\ref{fig:eval3_2} shows the MAP of the recommendation lists. We see that using co-download results in higher MAP than co-citation for the first two years after publication of a paper. Hence, to find related documents to recent papers, co-download is more informative. In this collection, two years after publication there are usually sufficiently many citations for co-citation to catch up.

While co-download performance slightly decreases beginning two years after publication, we believe that this is an artifact of the experiment design: the reference lists of papers citing older papers tend to contain other older papers. While co-citation benefits from this, co-download is penalized as it often recommends more recent related papers. Note that the experiment is further biased in favor of the co-citation measure, since (future) co-citations are used as ground truth.

Finally, the low absolute MAP values result from us considering only a single set of papers in the same references list as related for each evaluation. This means that when evaluating, there are often only about 10 other papers considered related out of the entire collection.

\section{Conclusions}

We have demonstrated that digital library access records are a valuable resource, containing implicit information about the relatedness of pairs of documents. 
We found that co-download is able to outperform citation-based recommendations on recently published papers in the arXiv collection. Furthermore, in contrast to co-citation, recommendations from co-download are available for practically all documents in the collection in a timely fashion and without need for expensive extraction and curation.
We conclude that access records can be used effectively in recommender systems for digital libraries, especially when citations are not available (e.g.\ for images, audio, video, documents without citations), where citation indexing is difficult, or where citations are rare or unlikely to be resolved.

We thank Paul Ginsparg and Simeon Warner (arXiv.org) for valuable discussions and for the data that enabled this research. We also thank Travis Brooks (SLAC/SPIRES). 
This research was supported by a Google gift and NSF CAREER Award IIS-0237381. The second author was partly supported by a Microsoft Research Fellowship.


\end{document}